\documentclass[oupdraft]{bio}

\usepackage{fixltx2e}
\usepackage{float}
\usepackage{breqn}

\usepackage{bm,graphicx,color,amsbsy,amsmath,placeins,
 amsthm,amsfonts,amssymb,setspace,lscape}
\usepackage{caption}
\usepackage{newfloat}
\DeclareFloatingEnvironment[name={Appendix Figure}]{suppfigure}

   \usepackage{mathtools}

\usepackage{epstopdf}

\def\bSig\mathbf{\Sigma}

\usepackage{amsfonts}
\usepackage{amsmath}
\usepackage[FIGTOPCAP]{subfigure}
\usepackage{url}

\newcommand{\PP}{\mathbb{P}}

\newcommand{\RR}{\mathbb{R}}

\newcommand{\bbeta}{\bm{\beta}}

\newcommand{\bkappa}{\bm{\kappa}}

\newcommand{\bnu}{\bm{\nu}}

\usepackage{circuitikz}

\allowdisplaybreaks

\begin{document}


\title{\vspace{-2em}
Estimating AutoAntibody Signatures to Detect Autoimmune Disease Patient Subsets}

\author{ZHENKE WU$^{\ast,1}$, LIVIA CASCIOLA-ROSEN$^2$, AMI A. SHAH$^2$, \\
ANTONY ROSEN$^2$, SCOTT L. ZEGER$^3$\\[4pt]
\textit{$^1$ Department of Biostatistics and Michigan Institute of Data Science, University of Michigan, Ann Arbor, Michigan 48109\\}
\vspace{0.6em}
\textit{$^2$ Division of Rheumatology, Department of Medicine, Johns Hopkins University School of Medicine, Baltimore, Maryland, 21224\\
\vspace{-0em}}
\textit{$^3$ Department of Biostatistics, Johns Hopkins University, Baltimore, MD 21205\vspace{-0em}}
\vspace{-0em}{$^*$zhenkewu@umich.edu}}

\markboth%
{Z. Wu and others}
{Estimating AutoAntibody Signatures to Detect Autoimmune Disease Patient Subsets}

\maketitle

\footnotetext{To whom correspondence should be addressed.}

\vspace{-1.5em}
\begin{abstract}
{ Autoimmune diseases are {characterized by highly specific immune responses against molecules in self-tissues. Different autoimmune diseases are characterized by distinct immune responses, making autoantibodies useful for diagnosis and prediction. 	In many diseases, the targets of autoantibodies are incompletely defined. Although the technologies for autoantibody discovery have advanced dramatically over the past decade, each of these techniques generates hundreds of possibilities, which are onerous and expensive to validate. We set out to establish a method to greatly simplify autoantibody discovery, using a pre-filtering step to define subgroups with similar specificities based on migration of radiolabeled, immunoprecipitated proteins on sodium dodecyl sulfate (SDS) gels and autoradiography [\textbf{G}el \textbf{E}lectrophoresis and band detection on \textbf{A}utoradiograms (GEA)].}  Human recognition of patterns is not optimal when the patterns are complex or scattered across many samples. Multiple sources of errors - including irrelevant intensity differences and warping of gels - have challenged automation of pattern discovery from  autoradiograms.

In this paper, we address these limitations using a Bayesian hierarchical model with shrinkage priors for pattern alignment and spatial dewarping. The Bayesian model combines information from multiple gel sets and corrects spatial warping for coherent estimation of autoantibody signatures defined by presence or absence of a grid of landmark proteins. We show the pre-processing creates more clearly separated clusters and improves the accuracy of autoantibody subset detection via hierarchical clustering. Finally,  we  demonstrate the utility of the proposed methods with  GEA  data  from  scleroderma patients. }
{Autoantibody signatures; Batch effect; Bayesian image registration; Clustering; Gel electrophoresis; Peak detection; Markov chain Monte Carlo; Measurement error; Scleroderma.
}
\end{abstract}

\vspace{-2.5em}
\section{Introduction}

Discovering disease subgroups that share distinct disease mechanisms is fundamental to disease prevention, monitoring and treatment. For example, in autoimmune diseases, {specific autoimmune responses are associated with distinct disease phenotypes and trajectories} \citep{rosen2016autoantigens}. {Defining the molecular markers of these subgroups has value, as these markers are of diagnostic and prognostic significance, and guide management and therapy. For example, an immune response to RNA polymerase III in scleroderma is associated with cancer; this immune response arises in response to a mutation in RNA polymerase III in that patient's cancer.  While many prominent specificities recognized by the immune response have been defined, many remain to be discovered. Although modern measurement technologies are revolutionizing the ability to define specificities, each technique results in hundreds of possibilities, which are onerous and expensive to validate. A simple technique identifies patterns of antibody reactivity based on} the abundance of different weighted antigens immunoprecipitated by patient sera. {Defining similar antibody reactivity patterns prior to applying one of the new discovery technologies would greatly simplify validation and therefore reduce the cost and improve the speed of antigen identification.}

To identify the autoantibodies present in a patient's serum, scientists {mix serum collected from each patient with radiolabeled lysates made from cultured cells.  These lysates contain a representation of all the proteins expressed in that cell type.  Antibodies in each patient’s serum recognize and bind tightly to the specific protein(s) in the lysate against which they are directed (termed “immunoprecipitation”). After further processing, electrophoresis is used to sort the immunoprecipitated} mixture of molecules using a crosslinked polymer or gel that separates the proteins by weight. Because different weighted molecules migrate with different speeds, the sorted molecules form distinct autoradiographed bands along the gel. By design, one gel can sort multiple samples on parallel lanes. Such experiments, referred to as gel electrophoresis autoradiography (GEA), serve to identify subsets that share one or more interesting observed bands. {It is noteworthy that the lysate proteins are present in their native conformation.  In our experience, many autoantibodies have epitopes that are conformationally dependent, giving GEA a powerful advantage over many of the new peptide-based (linear epitopes) sequencing technologies.} The method in this paper is designed to estimate a multivariate binary autoantibody signature for each sample that represents the presence or absence of autoantibodies over a grid of molecular weights, referred to as \textit{landmarks}.

To infer patient subsets, we  can cluster patients based upon the presence or absence of each band as well as other features of the radioactive intensities such as the peak scale and amplitude. There are two critical barriers to the successful implementation of this approach that we address. First, there are \textit{batch}, or \textit{gel effects} in the raw GEA data. By design, molecules of identical weight would migrate the same distance along the gel. This distance however varies by gel due to differential experimental conditions. Second, {gels are frequently slightly warped  as they electrophorese due to heating effects generated during the electrophoresis procedure and due to artifacts introduced during physical processing  of the gels.} As the size and complexity of GEA experiment database grow, the need for systematic, reproducible and scalable error correction has also grown. 

In this paper, we introduce and illustrate a novel statistical approach to pre-process the high-frequency GEA data which we show improve our ability to compare and cluster band patterns across samples. The pre-processing involves peak detection and batch effect corrections.  In particular, we propose a local scoring algorithm for peak detection that is computationally efficient and performs well for minor peaks (Section \ref{sec:peakdetection}). The detected peaks then enter the image alignment method that corrects batch effects in two steps: reference alignment and spatial dewarping. First, reference alignment calibrates multiple gels towards a common molecular standard. We perform piecewise linear stretching/compression by placing knots at the \textit{marker} or \textit{reference} bands present on all the images (Section \ref{sec:reference.alignment}). The reference-aligned gel images produce a set of peak locations that are then fitted by a novel hierarchical Bayesian model. The proposed model assumes that the smooth spatial gel deformations have deviated the observed peaks from their true landmarks. We use Markov chain Monte Carlo to estimate both the smooth warping functions and, for each detected peak, the posterior probabilities over a grid of landmarks where it is aligned. The Bayesian framework has the advantage of incorporating inherent uncertainty in assigning a peak to a molecular weight landmark. 

 { The \textit{aligned} high-frequency intensity data (Section \ref{sec::exact_matching}) may be the input of many methods including hierarchical clustering, latent class models and factor analyses. In this paper, we focus on illustrating the value of alignment for the standard hierarchical clustering applied to data with known and unknown clusters (Section \ref{sec:application}).}   At each iteration of the MCMC sampling, we obtain the multivariate binary signatures that represent autoantibody presence or absence over a grid of landmarks and align the gel images.  Upon hierarchically clustering the aligned intensities at each iteration, we obtain a collection of dendrograms. In particular, we use the standard correlation-distance based agglomerative hierarchical clustering to create nested subgroups. For $N$ samples, hierarchical clustering produces a dendrogram that represents a nested set of clusters. Depending on where the dendrogram is cut, between 1 and $N$ clusters result. We demonstrate through real data that pre-processing more clearly separates the estimated clusters and improves the accuracy of cluster detection compared to naive analyses done without alignment.

The rest of the paper is organized as follows. Section \ref{sec:preprocessing} introduces the importance of pre-processing GEA data followed by algorithmic details for peak detection in Section \ref{sec:peakdetection} and batch effect correction in Section \ref{sec:batch.effects}.  In Section \ref{sec:computing}, we describe model posterior inference by MCMC and the statistical property of the shrinkage priors. We demonstrate how the proposed methods function through an application to signature estimation and subgroup identification of scleroderma patients in Section \ref{sec:application}. The paper concludes with a discussion on model advantages and opportunities for extensions.

\section{Data Pre-Processing}
\label{sec:preprocessing}

\subsection{GEA Data and Pre-processing Overview}
\label{sec:gelexperiments}
Gel electophoresis for autoantibodies (GEA) is designed to separate {autoantibody} mixtures according to molecular weight and to radioactively map them as bands along the gel. Figure \ref{fig:raw_image} shows an example of a raw GEA image. We tested four sets of samples from scleroderma patients {with a malignancy; of note, these sera were pre-selected as being negative for the three most commonly found scleroderma autoantibodies (anti-topoisomerase 1, anti-centromere and anti-RNA polymerase III antibodies, which in aggregate are found in about $60\%$ of scleroderma patients).} 

Each sample set consisted of $19$ patient sera plus one reference. In the middle panel of Figure \ref{fig:raw_data}, seven red vertical lines indicate the reference molecules of known weight (200, 116, 97, 66, 45, 31, 21.5) kDa. It also shows the band patterns that read out autoantibodies present in each of $19$ patient samples (lanes 2-20). The top of Figure \ref{fig:raw_data} shows the intensities from all the lanes; Seven clear spikes above the vertical lines again correspond to the reference molecules. The bottom of Figure \ref{fig:raw_data} shows the piecewise linear interpolation of the location-to-weight function using the seven reference weights as knots. The weight of an arbitrary peak (``*") can then be read from this interpolation. For example, the protein actin  (about 42kDa) produced the peaks immediately to the right of the 45kDa reference (the fifth vertical line from the left). The misalignment of the actin peaks is caused by non-rigid image deformation (Section \ref{sec:bayesian.warping}).

Identical reference molecules fail to align (empty circles, bottom panel of Figure \ref{fig:raw_data}) across multiple gels because of variation in experimental conditions such as the strength of the electric field. We correct such misalignment by matching the reference peak locations across gels and then piecewise-linearly stretch or compress each gel using the reference peaks as knots. The technique is referred to as \textit{piecewise linear dewarping} and was first used in human motion alignment anchored at body joints \citep[e.g.,][]{uchida2001piecewise}.

The autoradiographic process is also vulnerable to smooth non-rigid gel deformation. This is most evident from the bands of actin, a ubiquitous protein of molecular weight 42 kDa, present in all lanes at around $0.43$ (middle panel of Figure \ref{fig:raw_data}). The bands form a smooth curve from the top to the bottom. The curvature represents the gel deformation since actin has identical weight and should appear at identical locations across the $19$ lanes. Without correction, this deformation interferes with accurate sample comparisons even on the same gel.  In Section \ref{sec:bayesian.warping}, we propose a Bayesian hierarchical image dewarping model with shrinkage priors to correct the deformation and align the actin peaks.

To establish notations, let $(\mathbf{t}^{0}, \mathbf{M}^{0})=\left\{\left(t^{0}_{b}, M^{0}_{gib}\right)\right\}$ represent the standardized, high-frequency GEA data, for bin $b=1, \ldots, B$ on lane $i=1,\ldots, N_g$ from gel $g=1, \ldots, G$.  Appendix S1 describes the standardization of raw data. Here $\mathbf{t}^{0}$ is a equi-spaced grid over the unit interval $[0,1]$, where $t^{0}_{gb}=b/B\in [0,1]$, $b=1,\ldots, B$. $M_{gib}^{0}$ is the radioactive intensity scanned at $t^{0}_{b}$ for lane $i=1,\ldots, N_g$, gel $g=1,\ldots, G$. Let $N = \sum_{g}N_g$ be the total sample size.

For the rest of this section, we take the high-frequency data $(\mathbf{t}^{0}, \mathbf{M}^{0})$ and map it to multivariate binary data $\mathbf{Y}$ on a coarser common grid across gels. In Section \ref{sec:peakdetection}, we propose a general method to transform an arbitrary high frequency, nearly continuous intensity data into raw peak locations. We first apply the peak detection algorithm to $(\mathbf{t}^{0}, \mathbf{M}^{0})$ and obtain the peak locations $\mathcal{P}^{0}$. In Section \ref{sec:reference.alignment} we use the reference peaks, a subset in $\mathcal{P}^{0}$ from the first lane on each gel, to process $(\mathbf{t}^{0}, \mathbf{M}^{0})$ into reference-aligned data $(\mathbf{t}^{\sf R}, \mathbf{M}^{\sf R})$. In Section \ref{sec:bayesian.warping}, we transform the peaks $\mathcal{P}$ detected from $(\mathbf{t}^{\sf R}, \mathbf{M}^{\sf R})$ to a joint posterior distribution of a $N$ by $L$ binary matrix $\mathbf{Y}$ that represents presence or absence of a peak over a grid of $L$ landmarks for all the $N$ samples ($L=100$ in our application). In Section \ref{sec::exact_matching}, we will process the reference-aligned high-frequency data $(\mathbf{t}^{\sf R}, \mathbf{M}^{\sf R})$ into $(\mathbf{t}, \mathbf{M})$ where the peaks appear at the landmarks indicated by the ones in $\mathbf{Y}$.

\subsection{Peak Detection}
\label{sec:peakdetection}

This section presents a general algorithm for detecting peaks from intensity data. We illustrate the algorithm by detecting peaks $\mathcal{P}^{0}$  from data $(\mathbf{t}^{0}, \mathbf{M}^{0})$. The peaks may appear with varying background intensities. Because the occurrence of a local maximum is thought to be more important than the background level in autoantibody signature estimation, we design the algorithm to be insensitive to the absolute intensity level.

We adopted the following peak detection algorithm:

\begin{enumerate}

\item[i.] \textit{Local Difference Scoring}. For each bin $b=1,\ldots, B$, lane $i=1,\ldots, N_g$ of gel $g=1,\ldots, G$, calculate the local difference score by comparing the intensity at bin $b$ to its left and right neighbors exactly $h$ bins away and to the local minimum for locations in between $(t-h,t+h)$ (truncated at $1$ or $B$ if $b$ is near the endpoints). That is, we calculate
\begin{align}
\textsf{score}_{gi}(b)  & =
{\sf sign}\left\{M^{0}_{gib}-M^{0}_{gi,\ell(b)}\right\}+ 
{\sf sign}\left\{M^{0}_{gib}-M^{0}_{gi,r(b)}\right\}+\nonumber\\
& ~~~~ {\sf sign}\left\{M^{0}_{gib}-\min_{\ell(b) \leq b' \leq r(b)}{M^{0}_{gib'}}-C_0\right\},\label{eq:localscoring}
\end{align}
where ${\sf sign(a)}=1,0,-1$ indicates positive, zero, or negative values; $\ell(b)=\max\{b-h,1\}$ and $r(b)=\min\{b+h,B\}$ denote the left and right neighbors $h(=10)$ bins away, and $C_0$ denotes the minimum peak elevation. The tunning parameter $h$ controls the locality of the peaks and $C_0$ controls the minimum peak magnitude. 

\item[ii.]  \textit{Peak Calling}. We look for the bins among peak candidates defined by $\{b\mid {\sf score}_{gi}(b)=3\}$ that maximize their respective local intensities (see  Appendix S2 for details and alternative peak calling methods). Let $\mathcal{P}^{0}_{gi}$ represent the collection of the peak locations for lane $i$ and gel $g$.

\end{enumerate}

\noindent \textbf{Remark 1:}  The score defined in (\ref{eq:localscoring}) depends only on the \textit{sign}s of the differences in local intensities. They can be computed in parallel across all the samples. A two-dimensional analogue has been used in astrophysics to find low grey-scale intensity galaxies from telescope images \citep{xu2016detection}. 

\subsection{Batch Effect Correction}
\label{sec:batch.effects}

\subsubsection{Reference Alignment via Piecewise Linear Dewarping.}
\label{sec:reference.alignment}

Molecules with identical weight do not appear exactly at the same location in each lane of a single gel due to gel deformation or across gels due to variations in experimental conditions. We first align the reference peaks $\mathcal{P}^{0}_{g1}$, $g=1, \ldots, G$ via piecewise linear dewarping to address the gel-to-gel variation \citep{uchida2001piecewise}. In our application, we used seven reference molecules of known weight $(200, 116, 97, 66, 45, 31, 21.5)$ kDa.

We first match the reference peaks $\mathcal{P}^{0}_{g1}$ on a query gel $g$ to the reference peaks $\mathcal{P}_{g_01}$ on the template gel $g_0$, and then use the matched reference peaks and the endpoints as knots to linearly stretch or compress the gels. Quadratic or higher-order dewarping is also possible, but we found linear dewarping performs sufficiently well for our data.  Appendix S3 gives the details of the algorithm. We denote the high frequency, reference aligned data by $(\mathbf{t}^{\sf R}, \mathbf{M}^{\sf R})=\{(t^{\sf R}_{gib}, M^{\sf R}_{gib})\}$. Applying the peak detection algorithm in Section \ref{sec:peakdetection} to this data, we collect all the detected peaks in $\mathcal{P}=\{\mathcal{P}_g, g=1,\ldots, G\}$ where $\mathcal{P}_g$ represents the peaks from gel $g$.

\subsubsection{Bayesian Image Dewarping to Correct Gel Deformation.}
\label{sec:bayesian.warping}

Another source of error during autoradiographic visualization is the non-rigid, spatial gel deformation. The middle panel of Figure \ref{fig:raw_data} shows one such example. It also reveals three analytical challenges to be addressed before obtaining meaningful results from an automatic disease subsetting algorithm. First, some proteins, e.g., actin, are detected on multiple gels and must be aligned. The blue asterisks that denote the detected peaks near $0.43$, form a smooth but non-linear curve from the top to the bottom of the gel. Second, fewer bands appear on the right half of the image, because {these smaller proteins tend to contain fewer methionine residues for radiolabeling}. Higher estimation uncertainty of the dewarping function is therefore expected for the right half. Third, the observed locations of the peaks are likely random around their true locations as the result of the multiple sources of error. 

To address these issues, we designed a hierarchical Bayesian dewarping algorithm for two-dimensional images. The algorithm simulates presence/absence data from the conditional distribution of protein occurrence over a grid of equi-spaced landmarks given the detected peaks $\mathcal{P}$ from the prior pre-processing. The stochastic model is defined on a coarser grid of landmark proteins, $\bnu=\{0=\nu_0 < \nu_1 < \ldots < \nu_L < \nu_{L+1}=1\}$ where $\nu_\ell = \ell/(L+1)$, $\ell=0, 1, \ldots, L+1$. In this paper, we will align peaks only to the internal knots $\{\nu_{\ell}, \ell\neq 0,L+1\}$; $\nu_0$ and $\nu_{L+1}$ will be used in the boundary constraint (\ref{eq:boundary_constraints}) to ensure endpoint alignment for all the sample lanes. We introduce a novel shrinkage prior to promote alignment of peaks to a common landmark. We also introduce shrinkage priors that regularize the overall smoothness of the spatial dewarping functions.

Let $(T_{gij}, u_{gij})$ denote the ({\sf location, lane number}) for peak $j=1, \ldots, J_{gi}$ on lane $i=1,\ldots, N_g$, gel $g=1, \ldots, G$. We fix $u_{gij}$ to take values in $\{1,2,\ldots, N_g\}$ and collect them in $\bm{u} = \{u_{gij}\}$ where $u_{gij}=u_{gi}$ if they belong to the same lane $i$. Let $P_g=\sum_i {J_{gi}}$ denote the total number of peaks on gel $g$ and $P = \sum P_g$. Let $\mathbf{T} = \{\bm{T}_g\}$, where $\bm{T}_g = (\ldots, T_{gi1},T_{gi2}, \ldots, T_{gi,{J_{gi}}}, \ldots)'$ collects the peak locations for gel $g=1, \ldots, G$. Both $\bm{u}$ and $\mathbf{T}$ are $P$-dimensional column vectors. For computational stability, without changing notation, we standardize $\mathbf{T}$, $\bm{u}$ and $\bm{\nu}$ by substract their means and dividing by their standard deviations. We now use $\mathcal{P}=\{\mathbf{T}, \bm{u}\}$ to denote the data for the Bayesian dewarping model.

\noindent \underline{Model Likelihood.} \textit{Peak-to-landmark indicators $\mathbf{Z}$.} Let $Z_{gij}$ take values in $\{1,\ldots,L\}$. For example, $Z_{gij}=3$ indicates that the $j$-th peak in lane $i$ on gel $g$ is aligned to landmark 3. Let $\mathbf{Z}=\{\bm{Z}_{g}, g=1, \ldots, G\}$ where $\bm{Z}_g=\{Z_{gij},j=1, \ldots, J_{gi}, i=1, \ldots, N_g\}$. Note that any $\mathbf{Z}$ can be converted to $N$ multivariate binary observations $\mathbf{Y} = \{(Y_{gi\ell}, \ell=1, \ldots, L)\}$ for the presence or absence of a landmark, where \({Y}_{gi\ell}=\mathbf{1}\left\{\ell \in \{{Z}_{gij}, j=1, \ldots, J_{gi}\}\right\}\), referred to as \textit{signature}. 

\vspace{1em}
\noindent \textit{Gaussian mixture model for aligning observed peaks $\mathbf{T}$.} We model $\mathbf{T}$ as observations from a Gaussian mixture model with $L$ components, each representing one landmark. Given $\mathbf{Z} = \{Z_{gij}\}$ and the spatial dewarping function $\mathcal{S}_g$ to be discussed later, we assume 
\begin{align}
p\left\{(\underbrace{T_{gij}=t}_{\substack{\text{peak}\\ \text{location} }}, \underbrace{u_{gi}}_{\substack{\text{lane}\\ \text{number} }})\mid \underbrace{Z_{gij}=\ell}_{\substack{\text{matched~to} \\
\text{landmark~}\ell}}, \underbrace{T_{gi,j-1}}_{\substack{\text{nearest~left} \\
\text{peak~location}}}, 
\underbrace{\mathcal{S}_g}_{\substack{\text{warping} \\
\text{function}}},
\underbrace{\sigma_{\epsilon}}_{\substack{\text{noise}\\ \text{level}}} \right\} = \begin{cases}
\phi\left(t; \mathcal{S}_g(\nu_\ell, u_{gi}), \sigma_{\epsilon}\right),&~t\in\mathcal{I}_{gij}(\nu_\ell, A_0);\\
0,&~\text{otherwise},
\end{cases} \label{eq:component_lkd}
\end{align}
$\ell=1,\ldots,L$, for peak $j=1, \ldots, J_{gi}$, lane $i=1, \ldots, N_g$, gel $g=1, \ldots, G$, where $\phi(\cdot; a,b)$ is the Gaussian density function with mean $a$ and standard deviation $b$, and $\mathcal{S}_g$ is an unknown smooth bivariate function that characterizes the deformation $(\nu, u) \mapsto \left(\mathcal{S}_g(\nu, u),u\right)$. 

\vspace{1em}
\noindent \textbf{{{Remark 2:}}} The peak location $T_{gij}$ is assumed to follow a Gaussian distribution with mean equal to $\nu_\ell$ plus a horizontal displacement $\mathcal{S}_g(\nu_\ell, u_{gi})$ and noise level equal to $\sigma_\epsilon$. We assume $\sigma_{\epsilon}$ is independent of landmark and lane. The density function (\ref{eq:component_lkd}) is positive only in the set $\mathcal{I}_{gij}(\nu_\ell, A_0)\overset{\Delta}{=}\left\{t: |t-\nu_\ell|< A_0 \text{~and~} t > T_{gi,j-1}\right\}$. The first inequality prohibits $T_{gij} $ being matched to distant landmarks and limits the search space for $Z_{gij}$ in our algorithm; the second inequality places order constraints on the observed peak locations $T_{gij}>T_{gi,j-1}, j=2, \ldots, J_{gi}-1$. We will restrict $Z_{gij}>Z_{gi,j-1}$ to avoid reverse dewarping.

%

\vspace{1em}
\noindent \textit{Bivariate smooth warping functions $\mathcal{S}_g$.} For gel $g$, we model the warping function $\mathcal{S}_g:\RR^2\rightarrow\RR$ using the tensor product basis expansion
\begin{eqnarray}
\mathcal{S}_g(\nu, u) & = &\sum_{s=1}^{T_\nu}\sum_{t=1}^{T_{u}}\beta_{gst}B_{g1s}(\nu)B_{g2t}(u),\label{eq:tensor_product}
\end{eqnarray}
where $B_{g1s}(\cdot)$ and $B_{g2t}(\cdot)$ are the $s$-th and $t$-th cubic B-spline basis with intercept, and $\bkappa_{\nu}$ and $\bkappa_{u}$ are the knots along the two coordinate directions, respectively \citep[][Chapter 5]{friedman2001elements} and $T_{\nu}$ and $T_{u}$ are the total number of bases. In subsequent analyses, we choose $\bm{\kappa}_\nu$ with $T_\nu-4$ internal knots at the $s/(T_\nu-3)$-th quantile of $\{T_{gij}\}$, $s=1, \ldots, T_\nu-4$ and similarly for $\bm{\kappa}_u$. Let the two sets of B-spline basis functions along  $\nu$- and $u-$direction be $\bm{B}_{g1}(\cdot)=\left(B_{g11}(\cdot), \ldots, B_{g1T_\nu}(\cdot)\right)'$ and $\bm{B}_{g2}(\cdot)=(B_{g21}(\cdot), \ldots, B_{g2T_u}(\cdot))'$, respectively. 

However, valid spatial gel deformations are limited to gel stretching, compression or shift along the $\nu$ direction. We thus constrain the shape of $\mathcal{S}_g, g=1, \ldots, G$ by
\vspace{-0.5em}
\begin{eqnarray}
\textit{Monotonicity}: &&\nu_{0}\leq \mathcal{S}_g(\nu, u)< \mathcal{S}_g(\nu', u)\leq \nu_{L+1}, \forall \nu< \nu', \forall u; \label{eq:monotonicity}\\
\textit{Boundary Constraint}:&& \mathcal{S}_g(\nu_0, u)=\nu_0, \mathcal{S}_g(\nu_{L+1}, u)=\nu_{L+1}. \label{eq:boundary_constraints}
\end{eqnarray}

\begin{figure}[p]
\centering
\includegraphics[width=\linewidth]{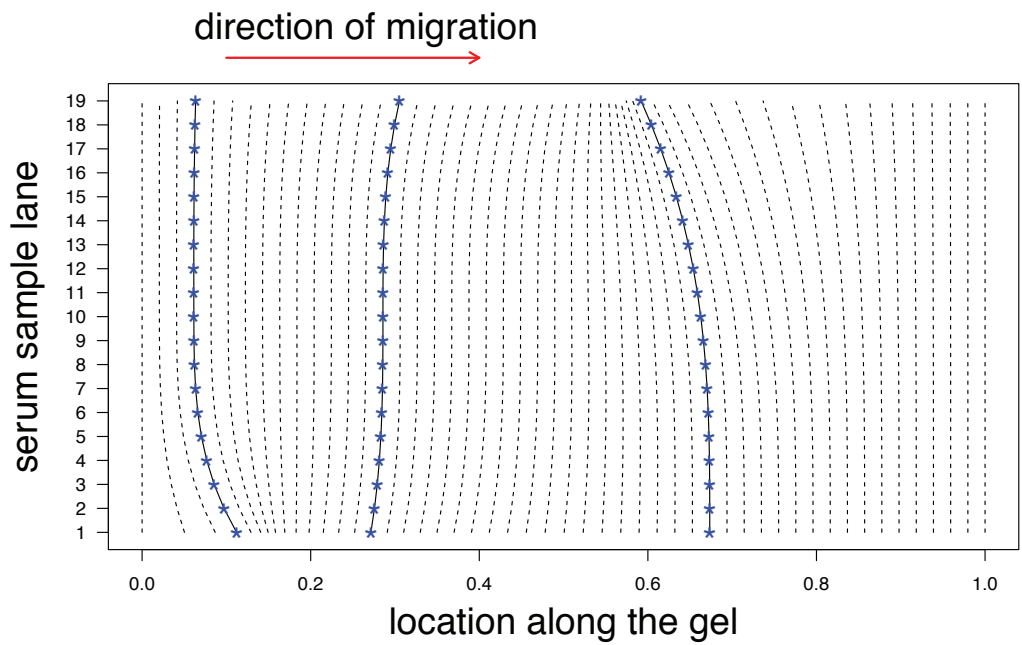}
\caption{The posterior mean estimate of a gel warping function $\mathcal{S}$ that corrects local stretching or compression. Highlighted are three vertical smooth curves, each of which aligns the peaks (blue asterisks ``{\color{blue} $\ast$}") with identical molecular weights.}
\label{fig:example_warping}
\end{figure}

\vspace{-0.5em}
The first constraint prevents reverse gel dewarping and the second assumes no gel shifting.  It can be relaxed to allow horizontal shifts by adding/substracting $\Delta$ for both equalities.  We implement both constraints by requiring the B-spline coefficients $\bbeta_g=\{\beta_{gst}\}$ to satisfy: $\nu_{0}= \beta_{g1 t}< \beta_{g2t}<\ldots<\beta_{g,T_{\nu}-1,t}< \beta_{gT_\nu t} = \nu_{L+1}$, $\forall t=1,\ldots,T_u$. Although only sufficient for $\mathcal{S}_g$'s monotonicity and boundary constraints, the foregoing $\bm{\beta}_g$ constraints allow flexible and realistic warpings. Figure \ref{fig:example_warping} shows a member warping function that corrects for local $``L"$-, $``S"$- and $``7"$-shaped deformations. 

The likelihood function (\ref{eq:component_lkd}) models the misaligned data $\mathcal{P}=\{\mathbf{T},\bm{u}\}$ in terms of the unknown spatial transformation $\mathcal{S}_g$ and the alignment $\mathbf{Z}$. Multiple raw gel images are then aligned by the model estimates accompanied by model-based uncertainty quantification. Importantly, coherent image registrations must align the universal actin peaks and hence require the borrowing of information among multiple misaligned observations. We accomplish this by sharing a set of intensity parameters $\{\lambda_\ell\}$ among the gels.

\noindent \underline{Prior}. \textit{Prior for} $\mathbf{Z}$. We describe a shrinkage prior for $\mathbf{Z}$ motivated by the need 1) to align the actin peaks (middle panel, Figure \ref{fig:raw_data}), and 2) to share the information about the location of actin peaks across multiple gels. 

We specify the prior distribution based on a discretized, non-homogeneous Poisson process with extreme intensities at a small number of landmarks.  Let the total number of the observed peaks follow a Poisson distribution: $J_{gi}  \overset{d}{\sim} {\sf Poisson}(\Lambda_g)$, for sample $i=1, \ldots, N_g$, gel $g=1, \ldots, G$. Given $J_{gi}$, let $Z^*_{gij} \mid \{\lambda_\ell^*\} \overset{iid}{\sim} {\sf Categorical}\left(\left\{ \lambda^*_\ell\right\}_{\ell=1}^L\right)$ describe which landmarks are present in sample $i$ of gel $g$. For sample lane $i$, we then define $\{Z_{gij}\}$ as the increasingly sorted $\{Z^*_{gij}\}$.  That is, we impose the order restriction $Z_{gij}\leq Z_{gij'}$ whenever peak $j$ appears to the left of peak $j'$ ($T_{gij}\leq T_{gij'}$). For hyperpriors, let $\lambda^*_\ell={\lambda_{\ell}}/{\sum_{\ell'=1}^L\lambda_{\ell'}}$ where $\lambda_\ell \mid \tau \overset{iid}{\sim} {\sf Normal}(0,\tau),  \ell=1,\ldots, L$, and the hyperparameter $\tau \overset{d}{\sim} {\sf Inv}$-${\sf Gamma}(10^{-4},10^{-4})$. Integrating over $\tau$, we obtain a marginal $t$-distribution for $\lambda_\ell$.

\vspace{0em}
\noindent \textbf{Remark 3:} It is easy to calculate the prior probability of landmark $\ell$ present in a sample $\PP(Y_{gi\ell}=1 \mid \lambda^*_\ell) \approx 1-\exp(-\lambda^*_\ell)$, $\ell=1, \ldots, L$, for large $L$. As shown by (A4) in Appendix, the ratio of the conditional posterior probabilities of assigning the peak $T_{gij}$ to landmark $\ell$ versus $\ell'$ is factorized into
\(\frac{\phi(T_{gij}; \mathcal{S}_g(\nu_{\ell}, u_{gi}),\sigma)}{\phi(T_{gij}; \mathcal{S}_g(\nu_{\ell'}, u_{gi}),\sigma)}\cdot\frac{1-\exp(-\lambda^*_\ell)}{1-\exp(-\lambda^*_{\ell'})}.\)
Suppose landmark $\ell$ is associated with a higher intensity, i.e., $\lambda^*_\ell>\lambda^*_{\ell'}$, the second ratio favors landmark $\ell$ given the likelihood ratio in the first term. Because the $\{\lambda^*_\ell\}$ are independent of $g$ and $i$, they globally modulate the probability of a landmark being present in all the gels. In our application, all the landmarks are \textit{in a priori} assumed to be equally likely by specifying independent $t$-distributed priors for the $\lambda_\ell$s. The $t$-distributions are heavy-tailed and can occasionally generate a large value of $\lambda_{\ell_0}$. Given $\lambda_{\ell_0}$, the posterior sampling algorithm will visit and then retain any configuration of $\mathbf{Z}$ that results in a large value of $\sum_{g,i}Y_{gi\ell_0}$ if the configuration substantially increases the joint posterior.

\vspace{1em}
\noindent \noindent \textit{Prior for} $\bm{\beta}_g$. 
We incorporate the prior knowledge that large and abrupt image deformations are rare.  We first specify priors for the horizontal basis coefficients $\beta_{gst}$, $s=2, \ldots, T_{\nu}-1$ at the $u$-direction basis $t=1$. We use a first-order random walk prior \citep{lang2004bayesian}
\begin{equation}
\beta_{gst} - \beta^{\sf\footnotesize id}_{s}\overset{d}{\sim} {N}\left(\cdot; \beta_{g,s-1,t}-\beta^{\sf\footnotesize id}_{s-1}, \sigma^{-2}_{g1}\right)\mathbf{1}\left\{\beta_{gst}\in (\beta_{g,s-1,t},\nu_{L+1})\right\}, t=1, s=2, \ldots, T_{\nu}-1,\label{eq:horizontal_randomwalk} 
\end{equation}
where $\bbeta^{\sf \footnotesize id}=(\beta^{\sf \footnotesize id}_1, \ldots, \beta^{\sf \footnotesize id}_{T_{\nu}})'$ is the vector of coefficients to represent an identity function $I: \nu\mapsto \nu$ in terms of the bases $\{B_{gs1}(\cdot)\}_{s=1}^{T_\nu}$; The truncation of $\beta_{gst}$ is needed for monotonicity (\ref{eq:monotonicity}). The hyperparameter $\sigma_{g1}^2$ controls the similarity between $\{\beta_{gs1}\}_{s=1}^{T_{\nu}}$ and $\bm{\beta}^{\sf id}$ and hence the similarity between $\mathcal{S}_g(\cdot, u)$ and the identity function $I$; $\sigma^2_{g1}=0$ represents no warping. We refer to $\sigma_{g1}^{-2}$ as the smoothing parameter along the $\nu$-direction.

Next, for any $s=2, \ldots, T_{\nu}-1$, we specify another random walk prior for the vertical basis coefficients
\begin{equation}
\beta_{gst} \overset{d}{\sim} {N}\left(\cdot; \beta_{gs,t-1},\sigma^{-2}_{gs}\right)\mathbf{1}{\{\beta_{gst}\in (\beta_{g,s-1,t},\nu_{L+1})\}}, t=2, \ldots, T_u.\label{eq:vertical_randomwalk}
\end{equation}
Similarly, the hyperparameter $\sigma^{-2}_{gs}$ controls the smoothness of $\mathcal{S}_g$ along the vertical or $u-$ direction; $\sigma^2_{gs}=0$ produces identical amounts of warping for all the lanes. Details about the hyperpriors for $\{\sigma_{gs}^2,s=1,2,\ldots, T_\nu-1\}$ are provided in  Appendix S4.

\label{sec:joint_distribution}
{\noindent {\it Joint Distribution.} The joint distribution of all the unknowns is
\begin{eqnarray}
&  & \prod_{g=1}^{G} \biggr\{ \underbrace{\prod_{i=1}^{N_g}\biggr[\prod_{j=1}^{J_{gi}} N\left(T_{gij}; \bm{B}_{g1}(\nu_{Z_{gij}})'\bm{\beta}_{g}\bm{B}_{g2}(u_{gi}), \sigma_{\epsilon}^{-2}\right)\mathbf{1}\{T_{gij}\in \mathcal{I}_{gij}(\nu_{Z_{gij}},A_0)\}}_{\sf likelihood~(\ref{eq:component_lkd})} \nonumber\\
&\times & \underbrace{J_{gi}!\prod_{j=1}^{J_{gi}}{\sf Categorical}(Z_{gij}; \bm{\lambda})\mathbf{1}\{Z_{gij}\leq Z_{gi,j+1},j=1,\ldots, J_{gi}-1\}\biggr]}_{\sf prior~of~\mathbf{Z}}
\nonumber \\
& \times & \underbrace{{N}_{T_{\nu}-1}\left(\{\beta_{gs1}\}_{s=1}^{T_{\nu}-1}; \bm{\beta}^{\sf id}_{[-T_{\nu}]}, \sigma_{g1}^{-2}\Delta_1'\Delta_1\right)\mathbf{1}\{\nu_0=\beta_{g11}<\ldots <\beta_{gs1} <\ldots < \beta_{g,T_\nu-1,1}<\nu_{L+1}\}\cdot p(\sigma_{g1}^2)}_{\sf prior~(\ref{eq:horizontal_randomwalk})~and~hyperprior~of~the~smoothing~parameter} \nonumber \\
& \times &\underbrace{ \prod_{s=2}^{T_{\nu}-1} \left[{N}_{T_u}\left(\bbeta_{gs\bullet}; \mathbf{0}, \sigma_{gs}^{-2}\Delta_2'\Delta_2\right) \mathbf{1}\{\nu_0=\beta_{g1t} < \beta_{g,s-1,t} <\beta_{gst} < \nu_{L+1}, \forall t \geq 2\}
\cdot 
p(\sigma^2_{gs}, \rho_{g})
\right]}_{\sf prior~(\ref{eq:vertical_randomwalk})~and~hyperpriors~of~the~smoothing~parameters}\biggr\}  {p(\bm{\lambda})},\label{eq:joint_distn}
\end{eqnarray}
where $p(\bm{\lambda})$, $p(\sigma_{g1}^2)$ and $p(\sigma^2_{gs}, \rho_{g})$ are the priors and hyperpriors and $N_d(\cdot; \bm{\mu}, \bm{\Lambda})$ denotes the $d$-dimensional multivariate normal density with mean $\bm{\mu}$ and precision matrix $\bm{\Lambda}$ (can be degenerate). The matrix $\Delta_1$ maps a column vector to its first-order differences (used in \ref{eq:horizontal_randomwalk}): $\Delta_{1kk'}=\delta(k+1,k')-\delta(k,k')$, $k=1, \ldots, T_{\nu}-2$, $k'=1, \ldots, T_\nu-1$, where $\delta(a,b)=1$ if $a=b$ and equals $0$ otherwise; Similarly we define $\Delta_2$ with $T_\nu$ replaced by $T_u+1$.
}

\section{Model Estimation and Implementation}
\label{sec:computing}

\subsection{Posterior Sampling}
\label{sec::posterior_sampling}
We use Markov chain Monte Carlo (MCMC) to simulate samples from the joint posterior distribution of all the unknowns \citep[e.g.,][]{gelfand1990sampling} and then draw posterior inferences about chosen functionals of the model parameters. Of special interest are the gel warping functions $\{\mathcal{S}_g(\cdot,\cdot; \bbeta)\}$ and the peak-to-landmark alignment $\mathbf{Z}$. Appendix S5 describes the sampling algorithm and discusses conditions for statistical identifiability of the warping functions. All the model estimation and visualization are performed by the \texttt{R} package \texttt{spotgear} (\url{https://github.com/zhenkewu/spotgear}).

Turning to dewarping a new GEA image $g^*$, we perform reference alignment (Section \ref{sec:reference.alignment}) and then obtain the peaks $\mathcal{P}_{g^*}$ (Section \ref{sec:peakdetection}). We approximate the joint posterior of $(\bm{\beta}_{g^*}, \bm{Z}_{g^*})$ by
\begin{eqnarray}
p(\bm{\beta}_{g^*}, \bm{Z}_{g^*} \mid \mathcal{P}, \mathcal{P}_{g^*}) & = 
& \int p(\bm{\beta}_{g^*}, \bm{Z}_{g^*} \mid \bm{\lambda}, \mathcal{P}_{g^*})p(\bm{\lambda} \mid \mathcal{P}, \mathcal{P}_{g^*}) \mathrm{d} \bm{\lambda}   \approx \int p(\bm{\beta}_{g^*}, \bm{Z}_{g^*} \mid \bm{\lambda},\mathcal{P}_{g^*})p(\bm{\lambda} \mid \mathcal{P}) \mathrm{d} \bm{\lambda},\nonumber
\end{eqnarray}
where the first term of the integrand is an one-sample conditional posterior and the second term is the posterior of $\bm{\lambda}$ given the old peaks $\mathcal{P}$.  Given $\mathcal{P}_{g^*}$, the first term can be derived from the joint distribution (\ref{eq:joint_distn}) with $G=1$. The integral can then be approximated by $K^{-1} \sum_k p(\bm{\beta}_{g^*}, \bm{Z}_{g^*} \mid \bm{\lambda}^{(k)},\mathcal{P}_{g^*})$ where $\{\bm{\lambda}^{(k)},k=1,\ldots, K\}$ are the stored posterior samples.

\subsection{Exact Peak Alignment for $\mathbf{M}$}
\label{sec::exact_matching}

We now describe the high-frequency data $(\mathbf{t}, \mathbf{M})=\left(\mathbf{t}, \mathbf{M}(\mathbf{Z})\right)$ that have peaks \textit{exactly} aligned according to $\mathbf{Z}$. We simply perform piecewise linear dewarping for each sample lane so that the detected peak $(T_{gij}, u_{gi})$ is horizontally adjusted to match its landmark $\nu_{{Z}_{gij}}$, $ j=1, \ldots, J_{gi}$. To do so, we apply the algorithm in Appendix S3 with $\{\nu_0, T_{gi1}, \ldots, T_{giJ_{gi}}, \nu_{L+1}\}$ as a query and $\left\{\nu_0, \nu_{{Z}_{gi1}}, \ldots, \nu_{Z_{giJ_{gi}}}, \nu_{L+1}\right\}$ as a template. In the following analyses, we will create aligned data using either 1) $\mathbf{Z}=\mathbf{Z}^{(k)}, k=1,\ldots, K$, the stored MCMC samples when calculating the posterior distributions of the parameters that are functions of $\mathbf{Z}$, or 2) $\mathbf{Z}=\widehat{\mathbf{Z}} :=\{\hat{{Z}}_{gij}= \arg \max_{\ell=1, \ldots, L} p(Z_{gij}=\ell \mid \mathcal{P})\}$, the \textit{maximum a posteriori} (MAP) alignment to obtain $\mathbf{M}(\hat{\mathbf{Z}})$.

\noindent {\bf Remark 4}. Because ${\mathcal{S}}_g$ is monotonic in $\nu$ given $u$, let ${\mathcal{S}}_g^{-1}(\cdot; u)$ denote its inverse.  One might be tempted to dewarp the images so that $(t,u)$ is horizontally aligned to $\left({\mathcal{S}_g^{-1}}(t; u), u\right)$. However, because a peak $T_{gij}$ varies around its mean $\mathcal{S}_{g}(\nu_{Z_{gij}})$, unless $\sigma_{\epsilon}^2=0$, the inverse mapping cannot guarantee that ${\mathcal{S}_g^{-1}}(T_{gij};u_{gi})$ is equal to $\nu_{{Z}_{gij}}$. 

\section{Applications to Scleroderma Patient Subsetting}
\label{sec:application}

Our methodology is motivated by the long-term clinical objective of finding an autoantibody signature that subsets autoimmune disease patients into groups with more homogeneous phenotypes and disease trajectories. The first step is to use the GEA data to cluster patients into subgroups with potential to have different outcomes. {We used sera from well-characterized patients with scleroderma and an associated cancer identified through the IRB-approved Johns Hopkins Scleroderma Center database \citep{Shah2017}.}                                                                                                                                                                                                                                                                                                                                                                                                                                                                                                                                                                                                                                                                                                To test our algorithms, we first analyze two GEA replicates each of 20 samples. Compared to the results of hierarchical clustering without pre-processing, we show our pre-processing method creates more clearly separated clusters. We also show our pre-processing improves the accuracy of cluster detection evaluated against the true matching. As a second test, we apply the pre-processing method  to GEA measurements on 76 patients with unknown clustering. We observe that the use of the pre-proecessing algorithm identifies clusters that are clearly separated and scientifically meaningful.

\subsection{Outline of Analyses}

\label{sec:analysis_outline}

For subsequent analyses, this section describes the steps of pre-processing, clustering and three metrics that evaluate the obtained clusters. 

\noindent \underline{\it Pre-processing}. We apply the peak detection algorithm in Section \ref{sec:peakdetection} followed by batch effect corrections as described in Section \ref{sec:batch.effects}. We exclude the reference lane on each gel when performing the two dimensional Bayesian dewarping.  We used $T_\nu=10$ and $T_u=6$ cubic B-spline basis functions in the horizontal and vertical directions, respectively. The dewarping functions are estimated by $\{\hat{\mathcal{S}}_g=\mathcal{S}_g(\cdot, \cdot; \hat{\bm{\beta}}_g)\}$ where $\hat{\bm{\beta}}_g$ is the posterior mean estimated by the empirical average of the MCMC samples. We also obtain the MAP estimate $\widehat{\mathbf{Z}}=\{\hat{Z}_{gij}\}$. The choice of the number of bases is crucial for the estimation of the warping functions \citep[e.g.,][]{lang2004bayesian}. For example, larger values of $(T_\nu, T_u)$ define a richer class of functions that can accommodate abrupt local image deformations. Visual inspection of the alignment of the actin peaks makes clear that more parsimonious models are preferred. Further improvements in knot selection is possible using knots on a nonequidistant grid so that more knots are placed where the spatial image warping is severe and the peaks are dense.

\noindent \underline{\it Clusterings}. Given the peak-to-landmark alignment $\mathbf{Z}$, we follow Section \ref{sec::exact_matching} to obtain peak-aligned images $\mathbf{M}=\mathbf{M}(\mathbf{Z})$ and then obtain clustering solutions. For example, let $\mathbf{M}=\mathbf{M}(\hat{\mathbf{Z}})$ where $\hat{\mathbf{Z}}$ is the MAP alignment. For each pair of sample $i$ and $i'$, we calculate the pairwise distances \(d(i,i') = 1-\widehat{\sf cor}(\mathbf{M}_{gi\cdot},\mathbf{M}_{gi'\cdot})\) where $\mathbf{M}_{gi\cdot} = (M_{gi1}, \ldots, M_{giB})'$ and $\widehat{\sf cor}(\cdot, \cdot)$ is the Pearson's correlation coefficient. Denote the $N$ by $N$ matrix of pairwise distances by $\hat{D}=\{d(i,i')\}$. We use $\hat{D}$ in the standard agglomerative hierarchical clustering with complete linkage to produce a dendrogram $\widehat{\mathcal{T}}={\mathcal{T}}(\hat{D})$. By varying the level of cutting the dendrogram $\widehat{\mathcal{T}}$, we obtain a nested set of clusterings $\hat{\mathcal{C}}(n)$, $n=2, \ldots, N$. We similarly denote the dendrogram produced without pre-processing by $\widehat{\mathcal{T}}^0 = {\mathcal{T}}(D^0)$ where $D^0$ is the correlation-based distance matrix computed from $\mathbf{M}^{0}$. We denote the nested clusters by  $\hat{\mathcal{C}}^{0}(n)$, $n=2, \ldots, N$. We will evaluate the obtained clusters by three criteria below.

\noindent \underline{\it Adjusted Rand Index}. We assess the agreement between two clusterings of the identical set of observations using the adjusted Rand index ({\sf aRI}; \citet{hubert1985comparing}). {\sf aRI} is defined by
\begin{eqnarray}
{\sf aRI}(\mathcal{C}, \mathcal{C}') & = & \frac{\sum_{r,c}{n_{rc}\choose 2}-\left[\sum_r {n_{r\cdot}\choose 2}\sum_c {n_{\cdot c}\choose 2}\right]/{N\choose 2}}{0.5\left[\sum_r {n_{r\cdot}\choose 2}+\sum_c {n_{\cdot c}\choose 2}\right]-\left[\sum_r {n_{r\cdot}\choose 2}\sum_c {n_{\cdot c}\choose 2}\right]/{N\choose 2}},
\end{eqnarray}
where $n_{rc}$ represents the number of observations                                                                                                                                                                                                                                                                                                                                                                                                                                                                                                                                                                                                                                                                                                            placed in the $r$th cluster of the first partition $\mathcal{C}$ and in the $c$th cluster of the second partition $\mathcal{C}'$, $\sum_{r,c}{n_{rc}\choose 2} (\leq 0.5\left[\sum_r {n_{r\cdot}\choose 2}+\sum_c {n_{\cdot c}\choose 2}\right])$ is the number of observation pairs placed in the same cluster in both partitions and $\sum_r {n_{r\cdot}\choose 2}$ and $\sum_c {n_{\cdot c}\choose 2}$ calculates the number of pairs placed in the same cluster for the first and the same cluster for second partition, respectively. {\sf aRI} is bounded between $-1$ and $1$ and corrects for chance agreement. It equals one for identical clusterings and is on average zero for two random partitions; larger values indicate better agreements between the two clustering methods.

\noindent \underline{\it Average silhouette}. We also evaluate the strength of each clustering method using the \textit{average silhouette} \citep{rousseeuw1987silhouettes}. For observation $i$, its silhouette $s(i)$ for a partition $\mathcal{C}$ compares the within- to the between-cluster average distances: $s(i)=[b(i)-a(i)]/\max\{a(i),b(i)\}$ where $a(i)$ is the average distance of $i$ to all other observations within the same cluster and $b(i)=\min_{C\in \mathcal{C}: i\notin C} \frac{\sum_{i'\in C} d(i,i')}{|C|}$ is the minimum average distance between $i$ and a cluster not containing $i$. $s(i)$ lies in $[-1,1]$ where a large value indicates observation $i$ is in a tight and isolated cluster. A larger \textit{average} silhouette $\bar{s}(\mathcal{C})=N^{-1}\sum_i s(i)$ indicates more clearly separated and tighter clustering $\mathcal{C}$.

\noindent \underline{\it Confidence levels of clusters}. In addition to the alignment uncertainty addressed by the posterior distribution $[\mathbf{Z}\mid \mathcal{P}]$, another source of uncertainty is the clustering of the aligned high-frequency intensity data $(\mathbf{t}, \mathbf{M}(\mathbf{Z}))$ given $\mathbf{Z}$. In this paper we chose not to specify the full probability distribution for the continuous intensities $(\mathbf{t}, \mathbf{M}(\mathbf{Z}))$. Following \citet{shimodaira2004approximately} and \citet{efron1996bootstrap}, we use bootstrap resampling to assess the confidence in the estimated dendrogram $\widehat{\mathcal{T}}$ (setting $\mathbf{Z}=\widehat{\mathbf{Z}}$, the \textit{MAP}). The bootstrap method perturbs the data by randomly sampling the columns of $(\mathbf{t}, \mathbf{M}(\widehat{\mathbf{Z}}))$ with replacement  and assesses the confidence levels for the presence of each subtree in $\widehat{\mathcal{T}}$. We calculate the frequency with which a subtree appears in an estimated dendrogram across all the bootstrap iterations where a large value (e.g., $> 0.95$) indicates strong evidence. We similarly bootstrap $(\mathbf{t}^0, \mathbf{M}^0)$ to assess the confidence in the dendrogram $\widehat{\mathcal{T}}^0$ estimated without alignment.

\subsection{Replication Experiments}
\label{sec::replicates}

Each of 20 biological samples were tested with two different lengths of exposure to autoradiographic devices: long (two-week) versus short (one-week) exposure. We ran 40 lanes on two gels that form 20 replicate pairs. Each gel image has 20 sample lanes: 19 serum sample lanes plus one reference lane comprised of molecules with known weights. The posterior dewarping results are shown in Appendix Figure S2.

We assess the agreement between the estimated clustering $\hat{\mathcal{C}}^{(k)}(n)$ and the true replication-based clusters $\mathcal{C}^*$ by {\sf aRI($\hat{\mathcal{C}}^{(k)}(n), \mathcal{C}^*$)}, for the number of clusters $n=2, \ldots, 20$ and the stored MCMC iteration $k=1, \ldots, K$. At iteration $k$, $\hat{\mathcal{C}}^{(k)}(n)$ is the clustering solution obtained by cutting the dendrogram that hierarchically clusters the peak-matched data $\mathbf{M}(\mathbf{Z}^{(k)})$ where $\mathbf{Z}^{(k)}$ is drawn from the posterior $[\mathbf{Z}\mid \mathcal{P}]$. 

\begin{figure}[p]
\centering     
\includegraphics[width=0.8\linewidth]{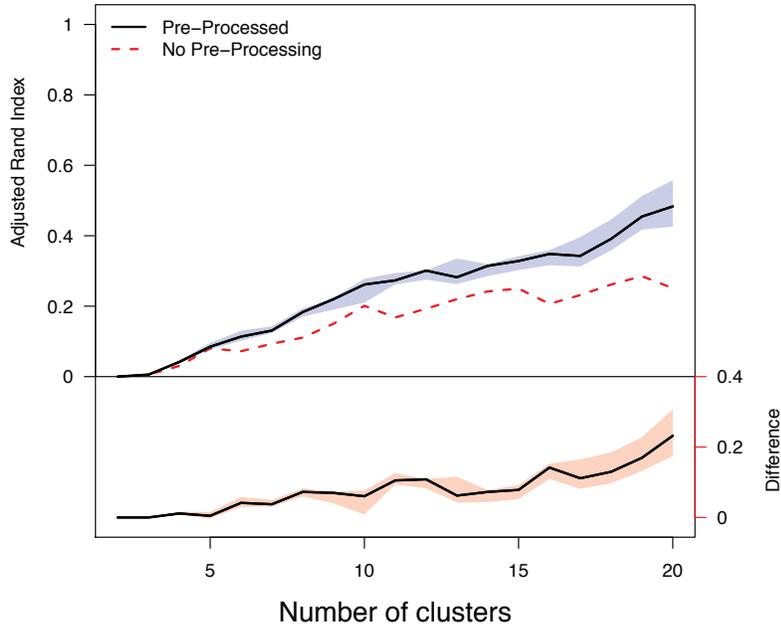}
\caption{Comparison between the adjusted Rand Indices obtained with and without pre-processing. \textit{Top}: The solid line and the blue shaded area represent the posterior mean {\sf aRI}s and the pointwise $95\%$ credible intervals, respectively. The dashed line is based on $\mathbf{M}^{\sf 0}$ without pre-processing. \textit{Bottom}: The solid line represents the difference between the {\sf aRI}s obtained with and without pre-processing: \{$K^{-1} \sum_{k}${\sf aRI($\hat{\mathcal{C}}^{(k)}(n), \mathcal{C}^*$)} $-$ {\sf aRI($\hat{\mathcal{C}}^{0}(n), \mathcal{C}^*$)}, $n=2, \ldots, 20$\}; the shared area shows the pointwise $95\%$  credible intervals.}\label{fig:replication_aRI}
\end{figure}

The pre-processing enhances the hierarchical clustering to produce clusters closer to the true replicate pairs. In Figure \ref{fig:replication_aRI}, the posterior mean of the adjusted Rand indices based on $K=5,000$ saved MCMC samples (solid line, $K^{-1}\sum_k$ {\sf aRI($\hat{\mathcal{C}}^{(k)}(n), \mathcal{C}^*$)}) are uniformly higher than the adjusted Rand indices based on data without pre-processing (dashed line, {\sf aRI}($\hat{\mathcal{C}^0}(n), \mathcal{C}^*$)). In the bottom panel, for every $n$, the posterior distribution for the difference between the two {aRI}s excludes zero increases with the numbers of clusters. In addition, the pairwise distances in $\hat{D}$ (obtained from $\mathbf{M}(\widehat{\mathbf{Z}})$) decreased by between $6.2$ and $66.4\%$ (mean $26.9\%$) relative to those in $D^0$. These decreases in the distances result in a dendrogram $\widehat{\mathcal{T}}$ that puts 13 replicate pairs at the terminal leaves as compared to 8 in  $\widehat{\mathcal{T}}^0$. 

We also observed uniformly increased confidence levels of the presence of true replicate pairs upon pre-processing. Appendix Figure S3 examines the confidence levels associated with each subtree with (hierarchically clustering the MAP-aligned data $\left(\mathbf{t}, \mathbf{M}(\widehat{\mathbf{Z}})\right)$) and without pre-processing. For example, for pair 18, the estimated confidence level increases from 0.71 to 1 after pre-processing; The confidence levels for detecting the pairs 2, 8 and 11 see similar increases from $0.67, 0.79, 0.66$ to $0.97, 0.86, 1$, respectively.  The increase in confidence levels is partly explained by the tighter clusters obtained after data pre-processing: the average silhouette computed from the MAP clustering, $\bar{s}\left(\hat{\mathcal{C}}(n)\right)$, increased $14.2-117.6\%$ ($0.03-0.18$ in magnitude) for $n=2, \ldots, 20$ clusters.

\subsection{Scleroderma GEA Data without Replicates}
\label{sec::second_data}

We ran 4 GEA gels, each with 19 patient sera and one reference lane. The sera are from scleroderma patients with cancer who are all negative for common autoantibodies to RNA polymerase III, topoisomerase I and centromere proteins. We had no other prior knowledge about known or novel autoantibodies at the time the study was conducted.  The sera were loaded in random order on each gel; the reference sample comprised of known molecules was always in the first lane. In the following, we describe the estimated dewarping, alignment and the resulting clusters.

\begin{landscape}
\begin{figure}[htp!]
\begin{center}
\includegraphics[width=\linewidth]{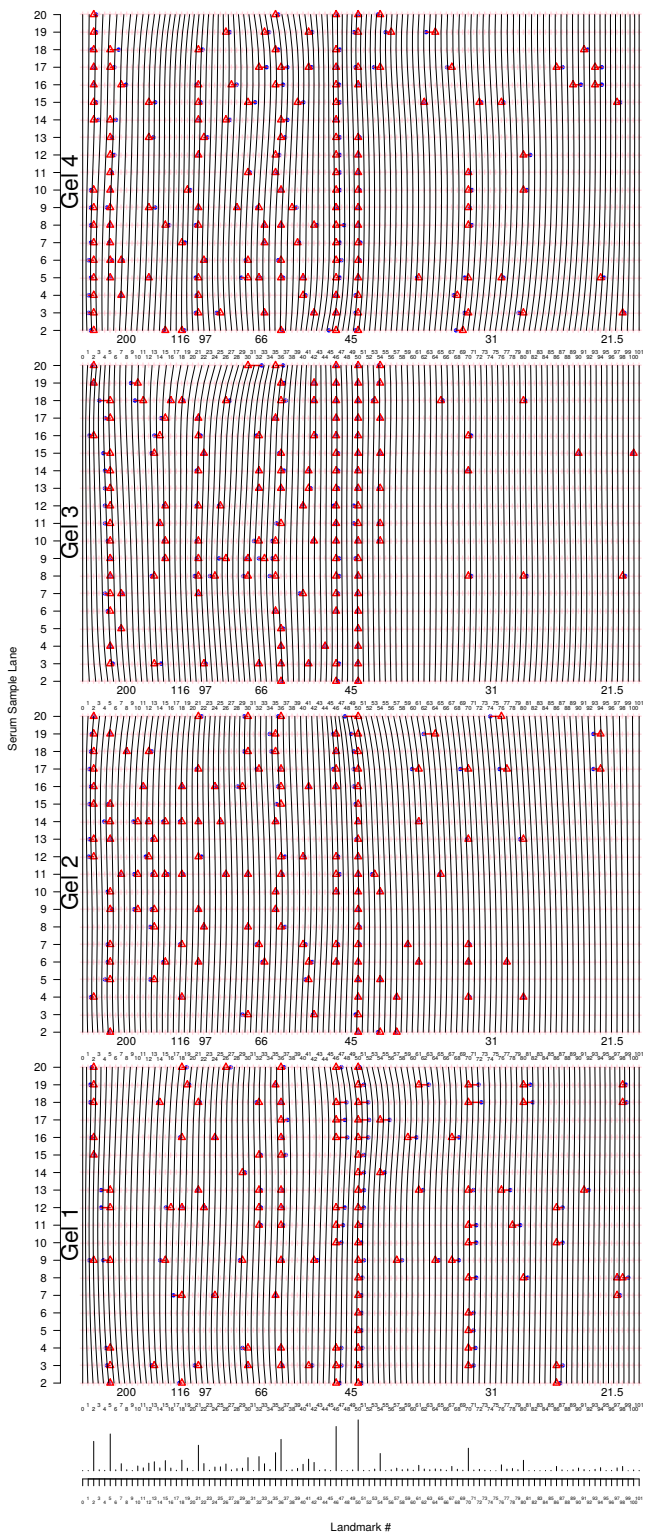}
\end{center}
\vspace{-0.1in}
\caption{Bayesian spatial dewarping results for the experiment without replicates. \textit{Top:} For each gel set, $19$ serum lanes over a grid of $L=100$ interior landmarks (reference lanes excluded). Each detected peak $T_{gij}$ (solid blue dots ``{\color{blue} $\bullet$}") is connected to its \textit{maximum a posteriori} landmark $\hat{Z}_{gij}$ (red triangle ``{\color{red}$\Delta$}"). The image deformations are shown by the bundle of black vertical curves $\{u \mapsto  \mathcal{S}_g(\nu_{\ell}, u): u = 2, \ldots, 20\}$, $\ell=1, \ldots, L, ~g=1, 2, 3, 4$, each of which connects the estimated locations of identical molecular weight. \textit{Bottom:} Marginal posterior probability of each landmark protein present in a sample.}
\label{fig:warped_all}
\end{figure}
\end{landscape}

\noindent \underline{\it Dewarping}. We pre-process the four gel sets by estimating the dewarping functions $\{\mathcal{S}_g, g=1, 2, 3, 4\}$ and the peak-to-landmark alignment $\mathbf{Z}$. We first removed a few spots on the right of the gels caused by localized gel contamination and assumed absence of peaks at these spots. The posterior dewarping results are shown in Figure \ref{fig:warped_all}. Each detected peak $\{T_{gij}\}$ (blue dot) is connected to its matched MAP landmark $\hat{Z}_{gij}$ (red triangle). The vertical bundle of black curves, one per landmark, visualizes the global shape of the estimated warping functions $\hat{\mathcal{S}}_g$. Along each estimated vertical curve, the locations $\left\{\left(\hat{\mathcal{S}}_g(\nu_\ell, u), u\right), \forall u\right\}$   represent identical molecular weights. 

\noindent \underline{\it Alignment to landmarks}. The marginal posterior probabilites of each landmark in a sample are shown at the bottom of Figure \ref{fig:warped_all}. For example, the posterior probability is $0.59$ for landmark 50 (about $43.4$ kDa, actin): the MAP estimate $\hat{\mathbf{Z}}$ shows that $73$ out of $76$ lanes. The marginal posterior probability is expected to further increase when more samples containing actin are analyzed via hierarchical Bayesian dewarping. Landmark 46 (about $46.6$ kDa) is another autoantibody hotspot where $54$ out of $76$ lanes have matched peaks. On the other hand, only $18$ and $1$ out of $76$ are matched to Landmarks 36 (about $59.8$ kDa) and 89 (about $23.4$ kDa), respectively. Their marginal posterior probabilities are hence low at $0.21$ and $0.01$. 

An animation of  the continuous dewarping process is available at \url{https://github.com/zhenkewu/spotgear}. It matches the detected peaks $T_{gij}$ to their MAP landmarks $\hat{Z}_{gij}$ and morphs the posterior mean dewarping $\hat{\mathcal{S}}_g$ into the constant function $\mathcal{I}: (\nu, u) \mapsto (\nu, u)$. Also shown is the pre-processed high-frequency data $(\mathbf{t}, \mathbf{M}(\hat{\mathbf{Z}}))$ with exactly matched peaks as described in Section \ref{sec::exact_matching}.

\begin{figure}[htp!]
\centering
\includegraphics[width=\linewidth]{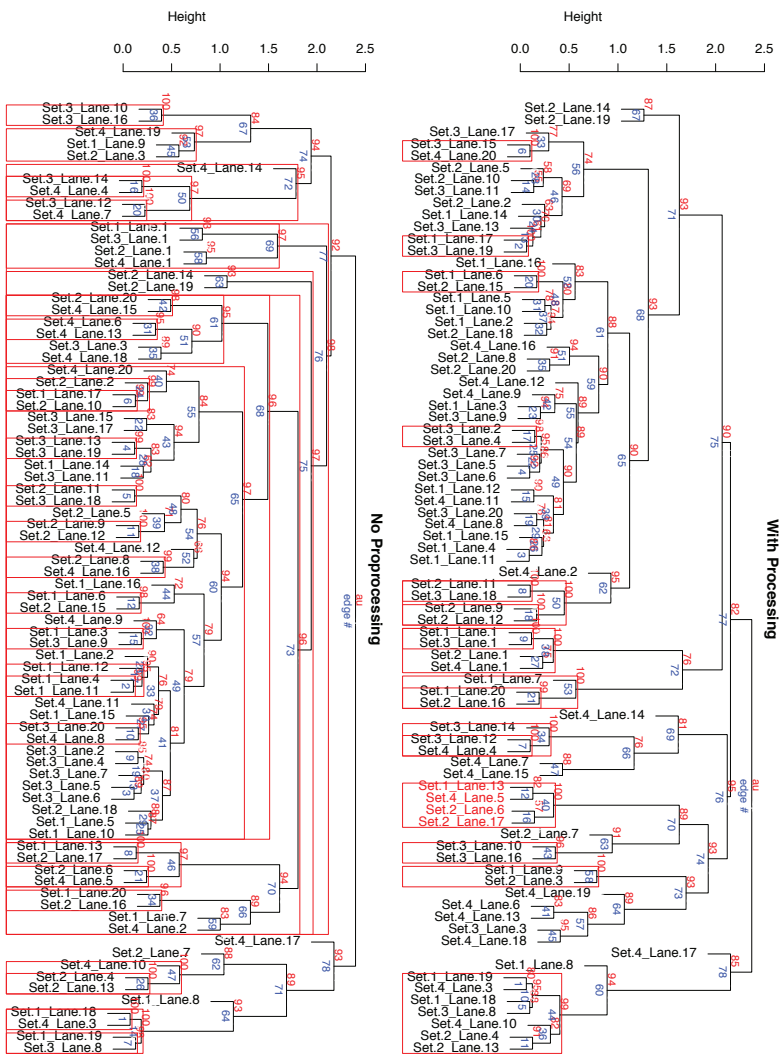}
\caption{Estimated dendrograms with (top) and without (bottom) pre-processing for the second data set. The red boxes show the subtree appearing in $>95\%$ of bootstrapped dendrograms with the actual estimated frequencies shown in red on top of the subtrees.
}
\label{fig:bootstrap_dendro_foursets}
\end{figure}

\noindent \underline{\it Clusters}. Our pre-processing method removed global warping phenomena and revealed a few strong clusters. The clusters with $0.95$ confidence levels or higher are shown in red boxes in Figure \ref{fig:bootstrap_dendro_foursets} for the analyses done with pre-processing (top) and without pre-processing (bottom). A comparison of the two clustering solutions favors the pre-processing approach. For example, within the dendrogram at the top, the first cluster from the right (number 44) consists of seven sample lanes ((Set, Lane): (1,19), (4,3), (1,18), (3,8), (4,10), (2,4), (2,13)) that are enriched at roughly $32.7$ and $27.9$ kDa. This group is split into two clusters (numbers 47 and 14) for the analyses done without pre-processing. In a second example, the cluster 46 at the bottom and cluster 40 at the top are comprised of identical samples (enriched at about $103.4$ kDa). We observe the confidence level increases from 0.97 to 1 after pre-processing. Pre-processing again produced more clearly separated clusters and eliminated many large clusters that are otherwise formed at the bottom of Figure \ref{fig:bootstrap_dendro_foursets}; We observed $8.8-39.5\%$ increases in the average silhouette based on the MAP alignment.

\section{Discussion}
\label{sec:discussion}

In this article, we have developed a novel statistical  approach to pre-processing and analyzing two-dimensional image data obtained from gel electrophoresis autoradiography (GEA). Our objective is to eliminate artifactual data patterns that can confound our ability to use standard clustering algorithms such as hierarchical clustering to detect subsets of autoimmune disease patients. The hierarchical Bayesian image dewarping model provides a natural framework for assessing uncertainty in the estimated alignment and warping functions and allows us to make inferences about many functions of parameters.

In Section \ref{sec:application}, we analyzed two sets of data from scleroderma patients. For the data with replication, we showed that the adjusted Rand indices increased if we perform pre-processing prior to standard hierarchical clustering. Based on the MAP alignment, the average silhouette that measures the strength of clustering increased by $14$ to $118\%$. The pre-processing also increased the confidence levels for detecting true replicates.

For the data without replicates, we showed that our pre-processing method successfully aligned the actin peaks. It also increased the confidence levels for the clusters that appeared in both clusterings (one with  pre-processing and the other without pre-processing). 

We conclude that there is added benefits of applying the pre-processing procedure prior to estimating disease subsets. We expect marginal though worthwhile gains to be achievable by using more carefully designed and tested tuning parameter selection procedure for local scoring (Section \ref{sec:peakdetection}). 

In the analysis of data with out replicates (Section \ref{sec::second_data}), we grouped the samples by creating a \textit{single} dendrogram given a fixed $\mathbf{Z}=\hat{\mathbf{Z}}$, i.e., the MAP alignment. Uncertainty exists in both the alignment and the dendrogram obtained by hierarchical clustering. We have addressed the former by the posterior distribution $[\mathbf{Z} \mid \mathcal{P}]$ and the latter by bootstraping. Future work is needed to assume a likelihood involving the unknown dendrogram structure to obtain and represent its posterior uncertainty \cite[e.g.,][]{chakerian2012computational}.

Two extensions based on prior biological knowledge are the current subject of further research. First, in our hierarchical Bayesian dewarping model, we assumed that the intensity parameters $\{\lambda^*_\ell\}$ are shared among the samples. However, the prevalence of autoantibodies may differ by subpopulation. For example, cancer versus non-cancer patients may have distinct distributions for the abundance of certain autoantibodies. We can either add another hierarchy on top of $\{\lambda_\ell^*\}$ or develop regression models for $\{\lambda^*_\ell\}$ to incorporate disease phenotype information and covariates such as age and gender. 

Second, proteins in the cells tend to work in complexes, so multiple autoantibodies are likely to be produced {against} a particular protein complex. This mechanism can be represented by a binary matrix $\mathbf{E}_{C\times L}$ where the $c$-th row $\left(E_{c1}, \ldots, E_{cL}\right)$ is a multivariate binary vector with $1$ for presence of landmark $\ell$ in complex $c$ and $0$ otherwise. The complexes are then assembled via $\bm{\eta}_{N\times L}=\mathbf{A}\mathbf{E}$ to produce the actual presence or absence of the landmarks for every patient, where $\mathbf{A}$ is a $N\times C$ binary matrix where each row represents the presence or absence of the $C$ complexes. Prior biological knowledge can be readily implemented via constraints on $\mathbf{A}$ or $\mathbf{E}$. For example, $A_{i1}=1$ for all the samples acknowledges the universal presence of autoantibodies produced against complex $1$, e.g., actin and likely others.  $\mathbf{A}$ and $\mathbf{E}$ can be inferred from the alignment indicators $\mathbf{Y}$ and continuous intensities. We may use regularization or shrinkage priors in a Bayesian framework to encourage a few maximally different complexes \citep[e.g.,][]{broderick2013}. One practical advantage of the Bayesian factorization approach lies in its convenient accommodation of repeated GEA on the same unknown sample by placing equality constraints on the rows of $\mathbf{A}$. Finally, our latent variable formulation $\bm{\eta}=\mathbf{A}\mathbf{E}$  makes it easy to incorporate multiple sources of patient lab and phenotype data that inform $\bm{\eta}$,  facilitate subgroup definition by $\mathbf{A}$ and perform individual predictions via the posterior predictive distributions of $\bm{\eta}$ \citep[e.g.,][]{BIOM:BIOM12577, wu2016partially, wu2017nplcm}.

\vspace{-3em}
\section*{Supplementary Materials}
Supplementary Material is available at
\url{http://biostatistics.oxfordjournals.org}.

\vspace{-2em}
\section*{Acknowledgments}
We thank the Editors, Associate Editor and two referees for their constructive suggestions that improved the presentation of our method and results. Research reported in this work was partially funded through a Patient-Centered Outcomes Research Institute (PCORI) Award (ME-1408-20318), NIH grant K23 AR061439 (to AS) and generous grants from the Jerome L. Greene Foundation and the Donald B. and Dorothy L. Stabler Foundation. The Johns Hopkins Rheumatic Disease Research Core Center, where the sera were processed and banked, and the antibody assays were performed, is supported by NIH grant P30 AR-070254.

\vspace{-2em}
\begingroup
    \setlength{\bibsep}{0.4em}
\bibliographystyle{biorefs}
\bibliography{gel}

\begin{thebibliography}{99}

\bibitem[Broderick \emph{and others}(2013)Broderick, Jordan and
  Pitman]{broderick2013}
\textsc{Broderick, Tamara, Jordan, Michael~I. and Pitman, Jim}. (2013, 08).
\newblock Cluster and feature modeling from combinatorial stochastic processes.
\newblock {\em Statist. Sci.\/}~\textbf{28}(3), 289--312.

\bibitem[Chakerian and Holmes(2012)Chakerian and
  Holmes]{chakerian2012computational}
\textsc{Chakerian, John and Holmes, Susan}. (2012).
\newblock Computational tools for evaluating phylogenetic and hierarchical
  clustering trees.
\newblock {\em Journal of Computational and Graphical
  Statistics\/}~\textbf{21}(3), 581--599.

\bibitem[Coley \emph{and others}(2016)Coley, Fisher, Mamawala, Carter, Pienta
  and Zeger]{BIOM:BIOM12577}
\textsc{Coley, Rebecca~Yates, Fisher, Aaron~J., Mamawala, Mufaddal, Carter,
  Herbert~Ballentine, Pienta, Kenneth~J. and Zeger, Scott~L.} (2016).
\newblock A bayesian hierarchical model for prediction of latent health states
  from multiple data sources with application to active surveillance of
  prostate cancer.
\newblock {\em Biometrics\/}, In Press.

\bibitem[Efron \emph{and others}(1996)Efron, Halloran and
  Holmes]{efron1996bootstrap}
\textsc{Efron, Bradley, Halloran, Elizabeth and Holmes, Susan}. (1996).
\newblock Bootstrap confidence levels for phylogenetic trees.
\newblock {\em Proceedings of the National Academy of
  Sciences\/}~\textbf{93}(23), 13429--13429.

\bibitem[Friedman \emph{and others}(2001)Friedman, Hastie and
  Tibshirani]{friedman2001elements}
\textsc{Friedman, Jerome, Hastie, Trevor and Tibshirani, Robert}. (2001).
\newblock {\em The elements of statistical learning\/}, Volume~1. Springer
  Series in Statistics Springer, Berlin.

\bibitem[Gelfand and Smith(1990)Gelfand and Smith]{gelfand1990sampling}
\textsc{Gelfand, Alan~E and Smith, Adrian~FM}. (1990).
\newblock Sampling-based approaches to calculating marginal densities.
\newblock {\em Journal of the American statistical
  association\/}~\textbf{85}(410), 398--409.

\bibitem[Hubert and Arabie(1985)Hubert and Arabie]{hubert1985comparing}
\textsc{Hubert, Lawrence and Arabie, Phipps}. (1985).
\newblock Comparing partitions.
\newblock {\em Journal of classification\/}~\textbf{2}(1), 193--218.

\bibitem[Lang and Brezger(2004)Lang and Brezger]{lang2004bayesian}
\textsc{Lang, Stefan and Brezger, Andreas}. (2004).
\newblock Bayesian p-splines.
\newblock {\em Journal of computational and graphical
  statistics\/}~\textbf{13}(1), 183--212.

\bibitem[Rosen and Casciola-Rosen(2016)Rosen and
  Casciola-Rosen]{rosen2016autoantigens}
\textsc{Rosen, Antony and Casciola-Rosen, Livia}. (2016).
\newblock Autoantigens as partners in initiation and propagation of autoimmune
  rheumatic diseases.
\newblock {\em Annual review of immunology\/}~\textbf{34}, 395--420.

\bibitem[Rousseeuw(1987)Rousseeuw]{rousseeuw1987silhouettes}
\textsc{Rousseeuw, Peter~J}. (1987).
\newblock Silhouettes: a graphical aid to the interpretation and validation of
  cluster analysis.
\newblock {\em Journal of computational and applied mathematics\/}~\textbf{20},
  53--65.

\bibitem[Shah \emph{and others}(2017)Shah, Xu, Rosen, Hummers, Wigley, Elledge
  and Casciola-Rosen]{Shah2017}
\textsc{Shah, Ami~A., Xu, George, Rosen, Antony, Hummers, Laura~K., Wigley,
  Fredrick~M., Elledge, Stephen~J. and Casciola-Rosen, Livia}. (2017).
\newblock Brief report: Anti¨crnpc-3 antibodies as a marker of
  cancer-associated scleroderma.
\newblock {\em Arthritis \& Rheumatology\/}~\textbf{69}(6), 1306--1312.

\bibitem[Shimodaira \emph{and others}(2004)Shimodaira
  et~al.]{shimodaira2004approximately}
\textsc{Shimodaira, Hidetoshi  \emph{and others}}. (2004).
\newblock Approximately unbiased tests of regions using multistep-multiscale
  bootstrap resampling.
\newblock {\em The Annals of Statistics\/}~\textbf{32}(6), 2616--2641.

\bibitem[Uchida and Sakoe(2001)Uchida and Sakoe]{uchida2001piecewise}
\textsc{Uchida, Seiichi and Sakoe, Hiroaki}. (2001).
\newblock Piecewise linear two-dimensional warping.
\newblock {\em Systems and Computers in Japan\/}~\textbf{32}(12), 1--9.

\bibitem[Wu \emph{and others}(2016)Wu, Deloria-Knoll, Hammitt and
  Zeger]{wu2016partially}
\textsc{Wu, Zhenke, Deloria-Knoll, Maria, Hammitt, Laura~L and Zeger, Scott~L}.
  (2016).
\newblock Partially latent class models for case--control studies of childhood
  pneumonia aetiology.
\newblock {\em Journal of the Royal Statistical Society: Series C (Applied
  Statistics)\/}~\textbf{65}(1), 97--114.

\bibitem[Wu \emph{and others}(2017)Wu, Deloria-Knoll and Zeger]{wu2017nplcm}
\textsc{Wu, Zhenke, Deloria-Knoll, Maria and Zeger, Scott~L.} (2017).
\newblock Nested partially latent class models for dependent binary data;
  estimating disease etiology.
\newblock {\em Biostatistics\/}~\textbf{18}(2), 200.

\bibitem[Xu \emph{and others}(2016)Xu, Postman, Meneghetti, Seitz, Zitrin,
  Merten, Maoz, Frye, Umetsu, Zheng  et~al.]{xu2016detection}
\textsc{Xu, Bingxiao, Postman, Marc, Meneghetti, Massimo, Seitz, Stella,
  Zitrin, Adi, Merten, Julian, Maoz, Dani, Frye, Brenda, Umetsu, Keiichi,
  Zheng, Wei  \emph{and others}}. (2016).
\newblock The detection and statistics of giant arcs behind clash clusters.
\newblock {\em The Astrophysical Journal\/}~\textbf{817}(2), 85.

\end{thebibliography}
\endgroup

\newpage

\begin{landscape}
\begin{figure}[p]
\captionsetup{width=0.8\linewidth}
\centering
\addtocounter{figure}{1} 
\raisebox{-.5in}{
\subfigure[]{
 \includegraphics[width=0.3\linewidth]
{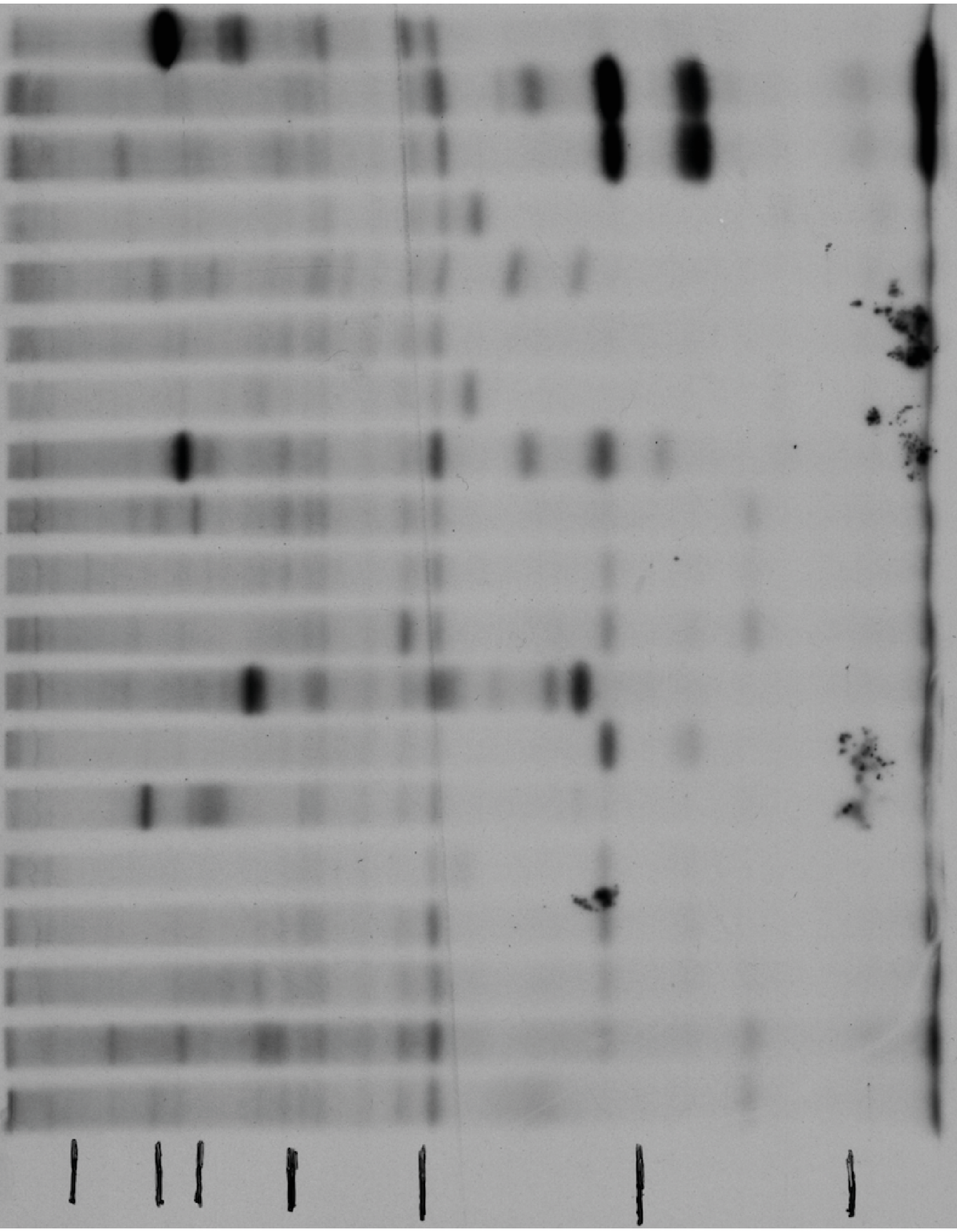}
\label{fig:raw_image}
}
}
\hspace*{.2in}
{\subfigure[]{
\includegraphics[width=0.6\linewidth]{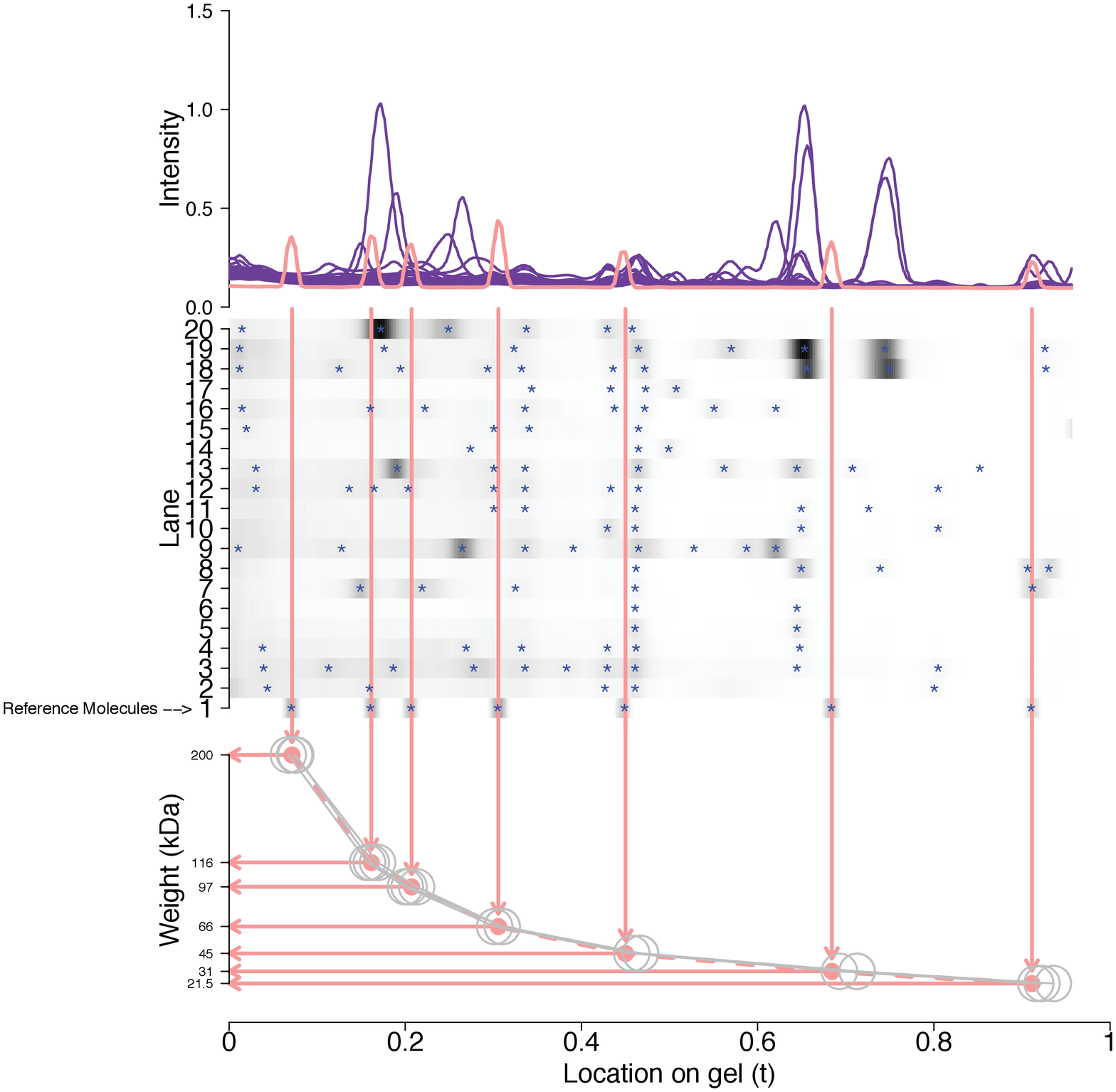}
\label{fig:raw_data}
}
}
\addtocounter{figure}{-1} 
\caption{Gel electrophoresis autoradiography data for $20$ samples on one gel. a) Raw GEA image. b) \textit{Top}: Radioactive intensities for all the samples; \textit{Middle}: Heatmap of the radioactive intensities for all the samples. The blue asterisks ({\color{blue} $\ast$}) denote the detected peaks. Seven vertical red lines indicate the locations of the seven reference molecules observed on lane 1. \textit{Bottom}: Actual molecular weights (Y-axis) as read from the location along the gel (X-axis). Four location-to-weight curves are shown here, each corresponding to reference lane 1s in the four gels analyzed in Section \ref{sec::second_data} (the dashed red curve ``- - -" is for the gel shown in the middle). Note the reference molecule misalignment shown by the scattered ``$\bigcirc$".}
\end{figure}
\end{landscape}


%
%

\setcounter{section}{0}
\setcounter{equation}{0}
\renewcommand{\theequation}{A\arabic{equation}}
\setcounter{figure}{0}
\renewcommand{\figurename}{}
\renewcommand{\tablename}{}
\renewcommand{\thefigure}{Figure S\arabic{figure}}
\renewcommand{\thetable}{Table S\arabic{table}}
\renewcommand{\thesection}{Appendix S\arabic{section}}
\renewcommand{\thesubsection}{\thesection.\arabic{subsection}}


\end{document}